\documentclass{iau} 
\usepackage{graphicx}

\title[Galaxy dynamics and intrinsic shapes]
{The intricate link between galaxy dynamics and intrinsic shape (or why so-called prolate rotation is a misnomer)}

\author[Caroline Foster \& Robert Bassett]   
{Caroline Foster$^{1,2}$,
 \and Robert Bassett$^{2,3}$}

\affiliation{$^1$ Sydney Institute for Astronomy, School of Physics, A28, The University of Sydney, NSW, 2006, Australia \\email: {\tt caroline.foster@sydney.edu.au}\\[\affilskip]
$^2$ ARC Centre of Excellence for All Sky Astrophysics in 3 Dimensions (ASTRO 3D) \\[\affilskip]
$^3$ Centre for Astrophysics and Supercomputing, Swinburne University
of Technology, PO Box 218, Hawthorn VIC 3122, Australia \\email: {\tt rbassett@swin.edu.au}}

\pubyear{2019}
\volume{353}  
\jname{Galactic Dynamics in the Era of Large Surveys}
\editors{J. Shen \& M. Valluri}
\begin{document}

\maketitle

\begin{abstract}
Many recent integral field spectroscopy (IFS) survey teams have used stellar kinematic maps combined with imaging to statistically infer the underlying distributions of galaxy intrinsic shapes. With now several IFS samples at our disposal, the method, which was originally proposed by M. Franx and collaborators in 1991, is gaining in popularity, having been so far applied to ATLAS$^{\rm{3D}}$, SAMI, MANGA and MASSIVE. We present results showing that a commonly assumed relationship between dynamical and intrinsic shape alignment does not hold in Illustris, affecting our ability to recover accurate intrinsic shape distributions. A further implication is that so-called ``prolate rotation'', where the bulk of stars in prolate galaxies are thought to rotate around the projected major axis, is a misnomer.
\end{abstract}

\firstsection 
\section{Introduction}

Projection effects hinder our ability to measure the distribution of three dimensional intrinsic galaxy shapes, leaving it a longstanding open question. One must rely on large samples to statistically reconstruct the intrinsic shape distribution of galaxies. For a review of the history of galaxy shape recovery studies, see \cite[M{\'e}ndez-Abreu (2016)]{MendezAbreu16}.

Galaxy shape is connected to a variety of galaxy properties such as age (\cite{vandeSande18}), spin (e.g. \cite{Foster17}), stellar mass (\cite{SanchezJanssen10}), luminosity (\cite{SanchezJanssen16}), morphology (e.g. \cite{Ryden06,Rodriguez13}), and environment (\cite{Fasano10,Rodriguez16}). Furthermore, theoretical simulations suggest that intrinsic shape depends on a galaxy's formation history (\cite{Jesseit09,Li18a,Li18b}). 

Analytical galaxy potentials suggest that there exists a simple relationship between galaxy intrinsic shape and the intrinsic misalignment angle between kinematic and morphological axes ($\Psi_{\rm int}$). Such a relationship reduces the number of unknowns, possibly providing more reliable inferred intrinsic shape distributions than methods using imaging alone. In  \cite[Bassett \& Foster (2019)]{Bassett19}, we explored the connection between galaxy intrinsic shape and stellar kinematics using cosmological hydrodynamical simulations from the Illustris project. This contribution summarizes one key result on the complex relationship between galaxy intrinsic shape and the intrinsic kinematic misalignment in Illustris.

\section{Simulations and method}\label{section:method}

We use data from the Illustris suite of hydrodynamic galaxy formation simulations (\cite{Vogelsberger14a}; \cite{Vogelsberger14b}; \cite{Genel14}). 
We select simulated galaxies from the $z=0$ snapshot of the largest Illustris simulation, Illustris 1, (see \cite{Vogelsberger14a} for details). Data are accessed through the Illustris public data release \cite[Nelson \etal\ (2015)]{Nelson15}. We select galaxies with $\geq10^{5}$ stellar particles to ensure robust intrinsic and projected shape measurements. This limit roughly corresponds to a stellar mass limit of $\sim10^{10}$ $M_{\odot}$ and is similar to that applied in other works on galaxy shape with Illustris (e.g. \cite{Li18b}). Our final sample contains 978 simulated galaxies.

\begin{figure}[b]
\begin{center}
 \includegraphics[width=3.4in, angle=270]{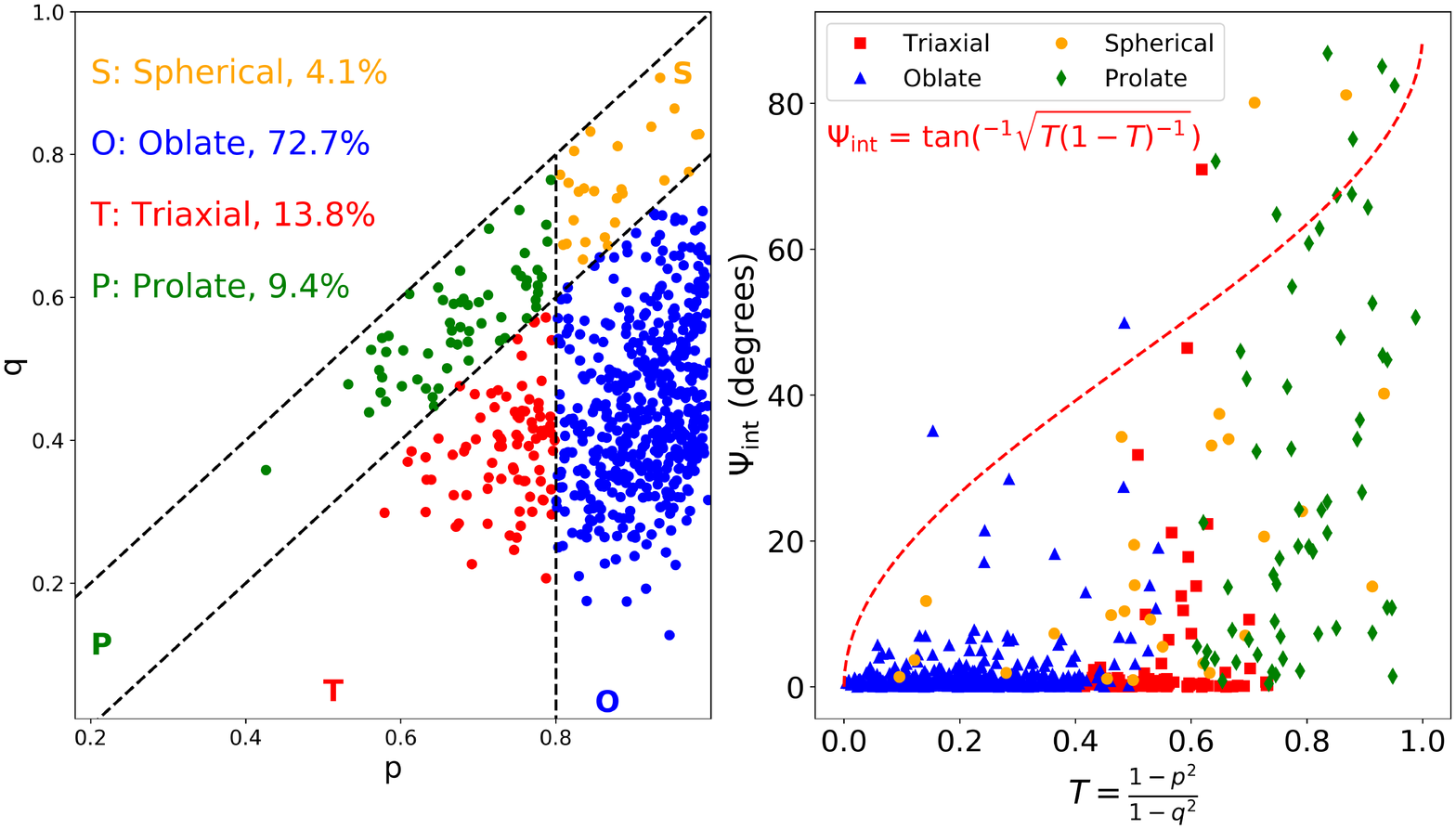} 
 \caption{LHS: Distribution of intrinsic axis ratios ($p=b/a$, $q=c/a$) for Illustris galaxies. Dashed lines delineate spherical, oblate, triaxial, and prolate as per \cite[Li \etal\ (2018)]{Li18b}. RHS: The distribution of triaxiality ($T$) vs intrinsic kinematic misalignment ($\Psi_{\rm int}$). Plotted symbols and colours indicate shape subclasses as labelled. The red dashed line represents the relationship suggested in \cite[Weijmans \etal\ (2014)]{Weijmans14}.}
   \label{fig1}
\end{center}
\end{figure}

\subsection{Measuring 3D Shape}\label{section:3dmeasure}

For each galaxy, we assume that the intrinsic shape of the distribution of stars can be approximated as a simple ellipsoid with three principal axes ($a \ge b \ge c$). We explicitly ignore the fact that galaxies may be composed of multiple components (e.g. bulge and disc) and instead consider their ``ellipsoid-equivalent'' shape. Under this assumption, the intrinsic shape of a galaxy is parametrised by two intrinsic axis ratios $p=b/a$ and $q=c/a$, where $0\le q\le p\le1$.

The square of the principal axes of the ellipsoid-equivalent shape, $a^2$, $b^2$ and $c^2$, correspond to the eigenvalues of the reduced inertia tensor ($I_{ij}$, similar to \cite{Allgood06,Li18b}), defined as:
\begin{equation}\label{eq:I}
	I_{ij} \equiv \sum_{n} \frac{x_{i,n}x_{j,n}}{\tilde{r}_{n}},
\end{equation}
where $\tilde{r}_{n}$, the 3D galactocentric distance of stellar particle $n$, is given by:
\begin{equation}
\tilde{r}_{n}=\sqrt{x_{n}^{2}+(y_{n}/p)^{2}+(z_{n}/q)^{2}},
\end{equation}
with the $x$, $y$, and $z$ axes aligned with the major, intermediate, and minor axes of the ellipsoid, respectively. The eigenvectors of the inertia tensor associated with the lowest and highest eigenvalue represent the minor and major axes of the ellipsoid-equivalent shape, respectively. The axis ratios $p$ and $q$ are then computed as the square root of the ratios of the corresponding eigenvalues. 

We employ an iterative process to compute the shape and select which particles to include because for galaxies produced in hydrodynamical simulations, reliable measurements of $p$ and $q$ require a careful exclusion of stellar particles at large radii where asymmetries can have a large impact on $I_{ij}$. We first select particles within the spherical half mass radius, $\tilde{r}_{e}$ and measure the galaxy shape following the above formalism. We then rotate the coordinate system to align the principal axes. Next, we use the latest $p$ and $q$ to update the half-mass radius $\tilde{r}_{e}$ and re-compute the eigenvalues of $I_{ij}$ using the updated selection of stellar particles. We repeat this process until $p$ and $q$ converge.

\subsection{Measuring the intrinsic kinematic misalignments}\label{section:2p2}

The intrinsic kinematic misalignment, $\Psi_{\rm int}$, is defined as the angle between the short axis of the equivalent ellipsoid and the stellar angular rotation vector, $\vec{R}$, in 3D. For consistency, $\vec{R}$ is measured for the same particles used to measure $p$ and $q$ above. $\vec{R}$ is defined as:
\begin{equation}
	\vec{R} \equiv \sum_{n} \vec{r_{n}} \times L_{n}\vec{v_{n}},
\end{equation}
where $L_{n}$ is the $r$-band luminosity of stellar particle $n$.

$\vec{R}$ is analogous to the angular momentum vector, where we have replaced the stellar mass by the $r$-band luminosity. We compute the intrinsic kinematic misalignment $\Psi_{\rm int}$ as:
\begin{equation}\label{eq:3dpsiint}
	\Psi_{\rm int} = 90^{\circ} - \cos^{-1}\left(\frac{\vec{R}\cdot\vec{e_{1}}}{|\vec{R}|}\right),
\end{equation}
where $\vec{e_{1}}$ is the unit vector parallel to the semi-major axis of the ellipsoidal-equivalent shape as defined in Section \ref{section:3dmeasure}. 
Note that in Equation \ref{eq:3dpsiint}, we define $\Psi_{\rm int}$ from the major axis rather than the minor axis because in near axisymmetric prolate galaxies ($p=q<1$), the orientation of the minor axis is ill-defined. This is less of an issue for axisymmetric oblate galaxies ($q<p$, $p=1$), which are almost exclusively aligned.

\subsection{Linking galaxy intrinsic shape to their kinematics}\label{section:2p2}

Following \cite[Li et al. (2018b)]{Li18b}, we divide galaxies into four basic shape groups based on their intrinsic axis ratios as shown in the left panel of Figure \ref{fig1}. Galaxies are classified as spherical, oblate, prolate, and triaxial as follows:\\
\indent\indent\indent Spherical: $\quad p-q < 0.2 \quad \& \quad p \geq 0.8$,\\
\indent\indent\indent Oblate: $\qquad p-q \geq 0.2 \quad \& \quad p \geq 0.8$,\\
\indent\indent\indent Prolate: $\quad \; \ \ p-q < 0.2 \quad \& \quad p < 0.8$,\\
\indent\indent\indent Triaxial: $\quad \; \  p-q \geq 0.2 \quad \& \quad p < 0.8$.\\
We compute the triaxiality parameter $T$ as per \cite[Franx, Illingworth \& de Zeeuw (1991)]{Franx91}: $T=\frac{1-p^2}{1-q^2}$.
Under this formalism, oblate ($a=b$) and prolate ($b=c$) systems have $T=0$ and $T=1$, respectively. Triaxial systems correspond to intermediate values of $T$ for which no two principal axes lengths are equal ($a\ne b\ne c$).

\section{Implications}

The right panel of Figure 1 clearly shows that Illustris galaxies do not follow the relationship between shape and $\Psi_{\rm int}$ put forward by \cite[Weijmans \etal\ (2014)]{Weijmans14}. The suggestion was based on theoretical work (e.g. \cite{Franx91}; \cite{Hunter92}; and \cite{Arnold94}) for elliptical galaxies with mass distributions described by St\"{a}ckel potentials, and underpins many observational efforts aimed at recovering the shape of galaxy samples from IFS observations (e.g. \cite{Foster17}).

Instead, Figure \ref{fig1} shows that the majority of oblate and triaxial Illustris galaxies are aligned (i.e. $\Psi_{\rm int}\simeq0$), while 
prolate and spherical galaxies exhibit a wide range in $\Psi_{\rm int}$. For spherical galaxies, this is partly due to the stochasticity of the direction of the major morphological axis. Prolate galaxies have a better defined major axis compared to spherical galaxies. For prolate galaxies, we find a mild excess of kinematically aligned ($\Psi_{\rm int} < 15^{\circ}$) galaxies, with the remaining galaxies uniformly scattering to higher values. Hence, even in the case of prolate galaxies with significant kinematic offsets there is no simple relationship. In particular, prolate galaxies do not have  $\Psi_{\rm int}\sim90^{\circ}$, which is often referred to as ``prolate rotation''. Prolate rotation is a misnomer: prolate galaxies need not rotate around their major morphological axis ($0\lesssim\Psi_{\rm int}\lesssim90^{\circ}$).
Unfortunately, $\Psi_{\rm int}$ holds little constraining power when inferring the distribution of galaxy intrinsic shapes. 

\bigskip
\noindent\emph{Acknowledgements:} The results presented herein were originally published in \cite[Bassett \& Foster (2019)]{Bassett2019}. This research was supported by the Australian Research Council Centre of Excellence for All Sky Astrophysics in 3 Dimensions (ASTRO 3D), through project number CE170100013.

{}

\end{document}